\documentclass[10pt,a4paper,notitlepage]{article}

\usepackage{amsmath,amssymb}
\usepackage[dvipdfm]{color}
\usepackage[dvipdfm]{graphicx,color}
\newcommand{\ket}[1]{\left| #1 \right>}
\newcommand{\bra}[1]{\left< #1 \right|}
\newcommand{\braket}[2]{\left< #1 \left.\!\! \right| #2 \right>}

\title
{Graph theoretical approach to evaluate molecular interactions in crystal.}

\author{Hiroya Nitta and Isao Kawata}

\begin{document}

\maketitle

\begin{abstract} 
We have applied the graph theory
 to estimation of interactions between electronic excitations in molecular crystals.
Intramolecular and intermolecular excitations have been considered.
The intermolecular interactions contain molecular integrals
 extended over two or three molecules,
 which we have called dimer or trimer.
%
The intermolecular interactions, such as $\pi-\pi$ stacking,
 is regarded as connections between neighboring molecules.
To take up all dimers and trimers in molecular crystal,
 we have treated the connections between molecules as edges of a graph.
The adjacency matrix of the graph and related quantity
 are utilized to count all dimers and trimers systematically.
The absorption spectrum of TiOPc crystal
 in which we have defined three types of intermolecular excitations is presented.

\end{abstract}

\section{Introduction}
Since the graph representation of molecule have been introduced by
 A. Crum Brown in chemistry \cite{Brown:JS8651800230},
 the graph theoretical approach is widely used in representing chemical structures.
Arther Cayley considered the number of isomers \cite{Cayley:10.1080/14786447408641058},
 then, molecular graphs have developed in mathematics.
Vertices and edges in molecular graphs correspond to
 atoms and bonds of a molecule, respectively.
Molecular graphs are utilized in order not only to sketch structures
of molecules, but to estimate their properties.
For example, the characteristic function of the adjacency matrix of a molecular graph
 is closely related to the Huckel Hamiltonian of the molecule.\cite{Huckel:1931}
Some topological invariants of molecular graphs are 
 important indices in quantitative sturucture property relationships (QSPR).\cite{Hosoya:19712332}
But the mainstream of quantum chemical calculation have developed
 without graph theoretical information of molecules.

The graph theoretical representation is also used in inorganic crystal structures.
Vertices correspond to atoms in crystal as well as the molecular sketch.
The crystals are build up by repetition of small unit cell
 which is constituted of several atoms,
 and bond network expands periodically,
 threfore, corresponding graphs have distinct freature from molecular graphs,
 that is infiniteness.
%
The correspoinding graph of an inoranic crsytal is infinite,
 but the connection of each atom to the neigbors (the degree of a vertex) is finite,
 therefore the graph is called locally finite.
From another point of view,
 it is known that by surjective map preserving local bond with neighbors
  the infinite crystal is mapped onto  finite graph.\cite{Sunada:noticesAMS}

Organic crystals have also periodic structure as inorganic ones,
 but interactions between molecules are weaker than covalent bonds.
In organic crystals, however,
 there are some important classes of intermolecular interactions, 
 e.g., $\pi$-$\pi$ interaction, CH-$\pi$ interaction, and intermolecular hydrogen bond.
W. D. S. Motherwell et al., utilized the graph theoretical approach
 in visualizing the hydrogen bonds in organic crystals.
 \cite{Motherwell:ha0182,Motherwell:ha0194}
The bonds are defined between hydrogen donor and acceptor,
 therefore, the edges are defined between atoms as well as
 normal molecular graphs and inorganic crystals.

Our subject is to estimate the interaction energies 
 including both intramolecular or local excitation (LE) and
 intermolecular charge transfer  excitation (CT)
 which is an electronic excitation from one molecule to another.
Because the coulomb interaction is not negligible even at long distance,
 all interactions between molecules in crystal have to be considered.
It is difficult that one estimates all electronic interactions in crystal
 because  the compuatational cost for quantum chemical calculation increases.
%
One idea to handle the problem is to classify the intermolecular interactions,
 and to apply practical approximation to every classified interaction.
%
%
In this study, we have applied the graph theory 
 to classify interactions and to count all molecular pairs interact with in crystal.
The CTs between molecules can be
 recognized as connections between neighbor molecules in crystal.
Therefore, we regard  molecules as vertices of a graph,
 and interacting molecular pairs   as edges.
In sec.\ref{sec:method}, we have summarized the method to calculate
 interaction energies between electronic excitations
 and counting problem in molecular crystals.
In sec.\ref{sec:results}, we have given a brief report of
 absorption spectrum of a molecular crystal.

\section{Method}\label{sec:method}
\subsection{Electronic excitations and molecular interactions in crystal}
Here, we consider spin singlet excitations in a molecule.
The electronic excitation from an occupied orbital $\psi_i$ of a molecule $A$
 to an unoccupied orbital $\psi_r$ of a molecule $B$
 is represented by a ket $\ket{\Psi_{i\rightarrow r}^{AB}}$.
If molecule $A$ and $B$ are same, this is LE type, otherwise CT type.
Based on configuration interaction method\cite{szabo1996modern},
 the general expression of interaction between excitations is
\begin{align}
\bra{\Psi_{i\rightarrow r}^{AB}} H \ket{\Psi_{j\rightarrow s}^{CD}}
&= \delta_{ij}\delta_{rs}\delta_{AC}\delta_{BD} (E_0 - \epsilon_i + \epsilon_r) \notag\\
& + \delta_{ij}\delta_{AC}(1-\delta_{BD})\bra{\psi_r^B} h \ket{\psi_s^D} 
 + \delta_{rs}\delta_{BD}(1-\delta_{AC})\bra{\psi_i^A} h \ket{\psi_j^C} \notag\\
& + \braket{\psi_i^A \psi_r^B}{\psi_j^C \psi_s^D}
 + \braket{\psi_i^A \psi_r^B}{\psi_s^D \psi_j^C},
\end{align}
where $\delta$  is the Kronecker's delta,
 $E_0$ is the ground state energy of the molecule,
 and $\epsilon_k$ is $k$-th orbital energy.
It is reasonable that the theory to evaluate the interactions between excited states
 is developed based on configuration interaction picture.
\cite{fujimoto:/content/aip/journal/jcp/137/3/10.1063/1.4733669}
The wavefunction-based methods evaluate correctly the coulomb interaction
 between two molecules which are spatially separated
 and not connected by chemical bonds.\cite{dreuw:2943}
The brackets in first three terms are expectation values of one-electron operator,
 and forth and fifth terms are those of two-electron operator.
The interactions we have been interested in especially, i.e.,
 the interaction between LEs (LE-LE),
 that between LE and CT  (LE-CT),
 and that between CTs (CT-CT), are
\begin{align}
\bra{\Psi_{i\rightarrow r}^{AA}} H \ket{\Psi_{j\rightarrow s}^{BB}}
&= \braket{\psi_i^A \psi_r^A}{\psi_j^B \psi_s^B}
 + \braket{\psi_i^A \psi_r^A}{\psi_s^B \psi_j^B}, \label{eq:LELEint} \\
\bra{\Psi_{i\rightarrow r}^{AA}} H \ket{\Psi_{j\rightarrow s}^{AB}}
&=\delta_{ij}\bra{\psi_r^A} h \ket{\psi_s^B} 
 + \braket{\psi_i^A \psi_r^A}{\psi_j^A \psi_s^B}
 + \braket{\psi_i^A \psi_r^A}{\psi_s^B \psi_j^A}, \label{eq:LECTint} \\
\bra{\Psi_{i\rightarrow r}^{AB}} H \ket{\Psi_{j\rightarrow s}^{AB}}
&=\delta_{ij}\delta_{rs} (E_0 - \epsilon_i + \epsilon_r)
 + \braket{\psi_i^A \psi_r^B}{\psi_j^A \psi_s^B}
 + \braket{\psi_i^A \psi_r^B}{\psi_s^B \psi_j^A}, \label{eq:CTCTdimer} \\
\bra{\Psi_{i\rightarrow r}^{AB}} H \ket{\Psi_{j\rightarrow s}^{AC}}
&=\delta_{ij}\bra{\psi_r^B} h \ket{\psi_s^C} 
 + \braket{\psi_i^A \psi_r^B}{\psi_j^A \psi_s^C}
 + \braket{\psi_i^A \psi_r^B}{\psi_s^C \psi_j^A}. \label{eq:CTCTtrimer}
\end{align}
The eq.\eqref{eq:LELEint} is the interaction energy between the LE of molecules $A$ and $B$.
The eq.\eqref{eq:LECTint} is the interaction energy 
 between the LE of molecule $A$ and the CT from molecules $A$ to $B$.
The eqs.\eqref{eq:CTCTdimer} and \eqref{eq:CTCTtrimer} are
 those between the CTs.
The eq.\eqref{eq:CTCTdimer} is that for the same molecular pairs,
 and the eq.\eqref{eq:CTCTtrimer} is that for different molecular pairs.
%
%
In eqs.\eqref{eq:LELEint}-\eqref{eq:CTCTtrimer},
 two-electron integrals are related to the dimer $\{A, B\}$,
 and  to the trimer $\{A,B,C\}$.
Therefore, we have been interested in dimers $\{\{A,B\},\{A,C\},\dots\}$,
and trimers $\{\{A,B,C\},\dots\}$.

To evaluate the excitaion spectrum of a molecular crystal with LE and CT, 
 we must collect all dimers and trimers in the crystal.

\subsection{Application of Graph theory}
We have applied the graph theory to handle the dimers and trimers.
In our definition,
 the vertices of the graph correspond to the molecules in crystal.
%
Such a graph can be represented by adjacency matrix $\mathbf{A}$,
\begin{align}
\mathbf{A} &= \begin{pmatrix}
A_{11} & A_{12} & \dots \\
A_{21} & &  \\
\vdots
\end{pmatrix},\\
\mathbf{A}_{AB} &= \begin{cases}
1 & \text{if molecule $A$ is connected to molecule $B$,} \\
0 & \text{otherwise.}
\end{cases} \notag
\end{align}
If molecules $A$ and $B$ interact, or CT from molecules $A$ to $B$ is considered,
 then $A$ and $B$ are defined as connected pair.
In our formalism, if ($A \rightarrow B$)  CT is considered,
 ($B \rightarrow A$) one is also done.
Therefore, the graph is non-directional, and the adjacency matrix is symmetric.
It is useful that two graphs for the LE-LE interaction
 and interactions related to CT are defined.
All dimers in crystal can be counted using the adjacency matrix $\mathbf{A}$.
%
%
Since our major subject is to calculate the interactions shown in eq.\eqref{eq:CTCTtrimer},
  all trimers $\{\{A,B,C\},\dots\}$ have to be taken up systematically
 based on the CT ralated graph.
In graph theory, the paths from one vertex to another through $l$ edges
  correspond to the elements of the matrix
\begin{align}
\mathbf{A^{(l)}} = \mathbf{A}^l.\label{eq:secpowerofA}
\end{align}
To collect the trimers, therefore, one can utilize the matrix $\mathbf{A^{(2)}}$.
If molecules A, B and C are components of a trimer, the condition
 ($\mathbf{A^{(2)}}_{BC} > 0$,
 and two or three of $\mathbf{A}_{AB}, \mathbf{A}_{BC}, \mathbf{A}_{AC}$ is 1)
 have to be fulfilled.
%
Higher order will be helpfull to estimate
 interactions dealt with four or more molecules.


\subsection{Construction of graph}
Because the coulomb interaction is crucial even at long distance,
 all intermolecular interactions in crystal have to be considered.
Considering a crystal which contain $M$ molecules,
 the number of interactions related to $N$ molecules is $_M \mathrm{C}_N$
 ($N=2$ for LE-LE, $N=2,3$ for LE-CT, and $N=2,3,4$ for CT-CT).
Such combinatorial number can be huge when $M$ is large,
 and the computational cost diverge.
To confront the problem,
 we classify the molecular interaction into two parts,
 and two graphs are defined for them.

First, let us consider the LE-LE interaction.
We have used multipole expansion of coulomb interaction to estimate LE-LE interaction.
The point-dipole approximation is often used  for LE-LE interaction.
\cite{0038-5670-7-2-R01,
doi:10.1021/ar50021a002,
doi:10.1021/ja00221a079,
doi:10.1021/j100099a037,
doi:10.1021/jp063879z}
Though we also used this approximation,
 higher order terms in the expansion can be taken in.
At the point-dipole approximation level,
 it is easy to estimate all interactions in the given crystal.
Such interactions are represented by the complete graph.
All off-diagonale elements of adjacency matrix for the graph are one.
For higher order expansion,
 it is reasonable that intermolcular distance of a pair  is limited, 
 and then other graphs and adjacency matrices will be defined.

Second, the graph of interactions related to CT are constructed as follows.
To estimate CT interaction is generally more difficult than LE,
 because CT interaction contain many-center molecular integrals essentially,
 which are bottle-neck of quantum chemical calculations.
It is known that the magnitude of CT interaction energy
 decreases with intermolecular distance increasing.
To limit the number of molecular pairs based on their distance
 is straightforward approach.
Another nature of CT is  directionality.
The main factor of CT in organic crystals are $\pi$ orbital overlap between molecules.
We construct the CT ralated graph based on both distance and orbital overlap.
Therefore, the graph and adjacency matrix for CT
 depend on the geometrical feature and molecular morphology of the given crystal.
Accuracy of the estimation depends mainly on intermolecular distance.
Above methods have been implemented to our lab made program.
Once we have collected all dimers and trimers in the given crystal,
 Hamiltonian including  CT and LE  have been constructed, 
 and the eigenvalues and eigenvectors of the Hamiltonian matrix have been calculated.


\section{Results and Discussion}\label{sec:results}
\begin{figure}
\includegraphics[clip,width=10cm]{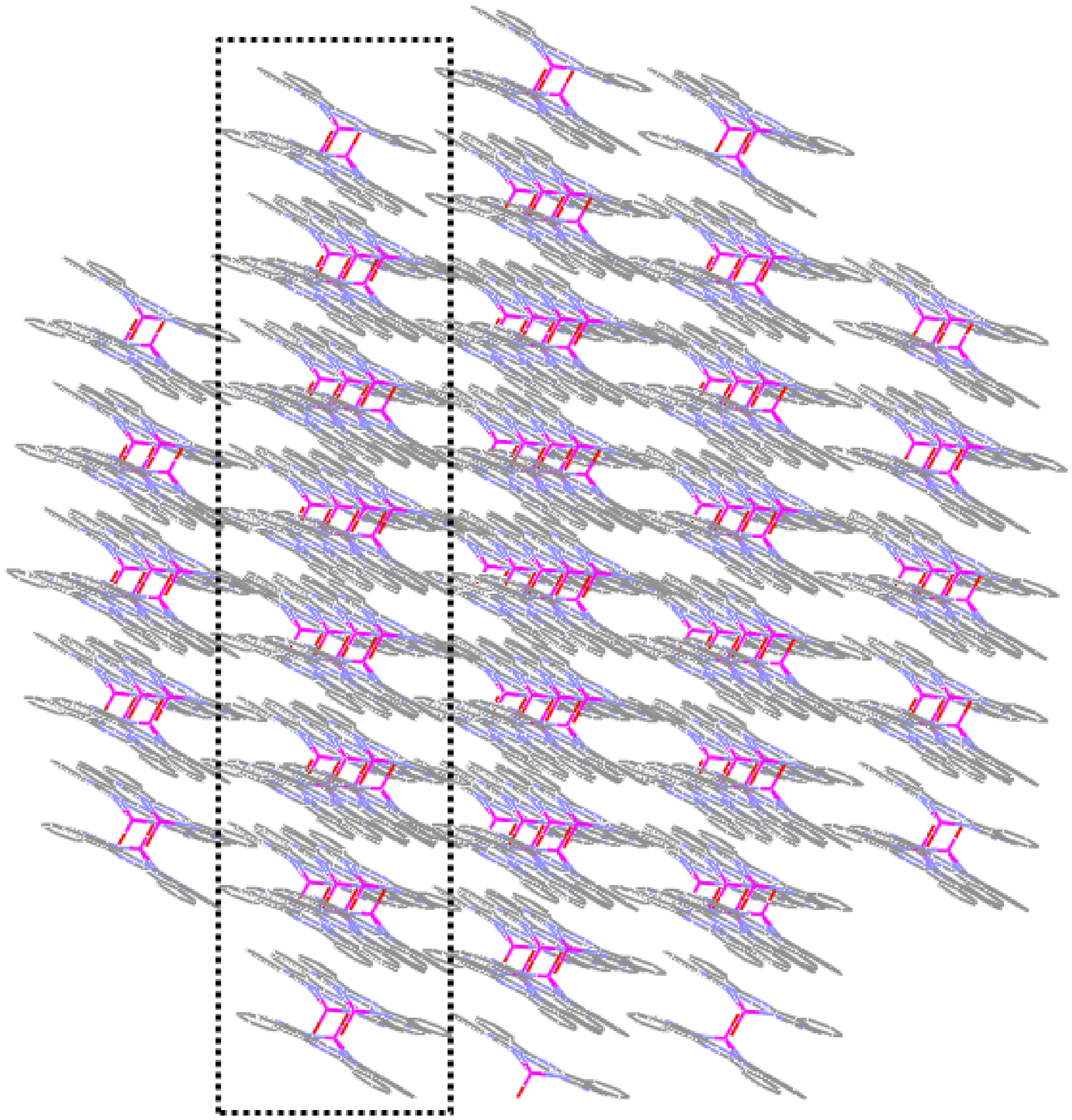}
\caption{A particle of TiOPc crystal (phase II) is shown. The particle is consist of 193 molecules. The CTs are defined in the network shown as dotted box.}
\label{fig:TiOPcII_crystal_193mol_network_clip}
\end{figure}

We have applied the methods in previous section
 to  evaluate the molecular interaction energies and
 to calculate molecular crystal absorption spectrum with LE and CT.
As a test, a spherical crystal, the radius of which is about 30 angstroms,
 is considered (Fig.\ref{fig:TiOPcII_crystal_193mol_network_clip}).
The molecule chosen here is known as titanyl phthalocyanine (TiOPc).
TiOPc crystals have several polymorphs,
 i.e., phase I, II, Y and so on.\cite{Hiller:tiopcpolymorph}
Here, we report the excitation spectrum of phase II (also called phase $\alpha$).

The LE of TiOPc is observed at about 700 nm 
 wavelength in solution.\cite{Zhang2010232,JM9960600143}
It is considered that
 interactions between transition dipoles of LEs of separated molecules
 are relatively strong even at long intermolecular distance.
At the point-dipole approximation,
 the interaction depends on $1/r^3$,
 where $r$ is the intermolecular distance.
We have constructed the graph for LE-LE interactions,
 and the edges are defined all considerable pairs ($_M \mathrm{C}_2$) in the crystal.

\begin{figure}
\includegraphics[clip,width=10cm]{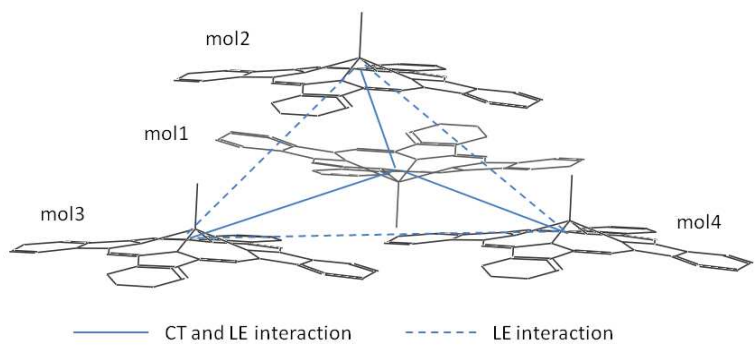}
\caption{Charge transfer pairs in TiOPc phase II crystal are shown.}
\label{fig:CTpairs_molnum_clip}
\end{figure}

\begin{figure}
\includegraphics[clip,width=10cm]{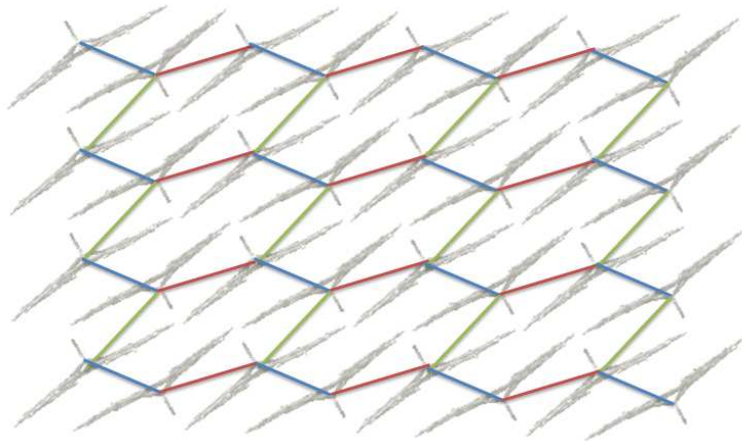}
\caption{Hexagonal lattice in TiOPc phase II crystal.}
\label{fig:hexlattice}
\end{figure}

Next, we consider the  graph for the interactions related to CT.
In our definition, the pairs not only adjoin in space
 but also stack along  $\pi$ orbital directions.\cite{Nakai:10.1021/jp035533j}
The pairs of CT and LE-LE interactions
 we have defined are depicted in Fig.\ref{fig:CTpairs_molnum_clip}.
The LE-LE interactions are defined between all molecular pairs,
 while the  interactions related to CT are defined only between 
 mol1 and mol2, mol1 and mol3, and mol1 and mol4.
It is noted that
 the CTs from mol1 to mol3 or vice versa are not defined in this study
 because orbital overlap between mol1 and mol3 is smaller than
 those of other molecular pairs.
The connections of corresponding three dimers are regarded as the basic edges
 of the graph for  interactions related to CT.
When expand the edges by translation, the connections form networks in the crystal.
One network is planar as depicted
 by dotted box in Fig.\ref{fig:TiOPcII_crystal_193mol_network_clip},
 and the crystal has layered structure of the networks.
The degree of each vertex is three as shown in Fig.\ref{fig:hexlattice}.
The network is known as hexagonal lattice,
 and the well-known example of the lattice is graphite.
Although graphite is the higher symmetric realization of the hexagonal lattice,
 the standard realization\cite{Sunada:noticesAMS},
 it is noted that CTs in TiOPc phase II crystal form a realization of the lattice.

Then, we have analyzed the type of interactions of pairs.
We have considered three types of CT pair 
 in this study(Fig.\ref{fig:CTpairs_molnum_clip}).
In the figure, molecules mol2, mol3 and mol4 are crystallographically identical,
 however, the interactions between mol1 and mol2, mol1 and mol3,
 and mol1 and mol4 are different.
If we have defined the weighted graph based on the interaction pattern,
 it will be usefull to reduce the computational cost
 to estimate the  interactions related to CT.

\begin{figure}
\includegraphics[clip,width=10cm,angle=270]{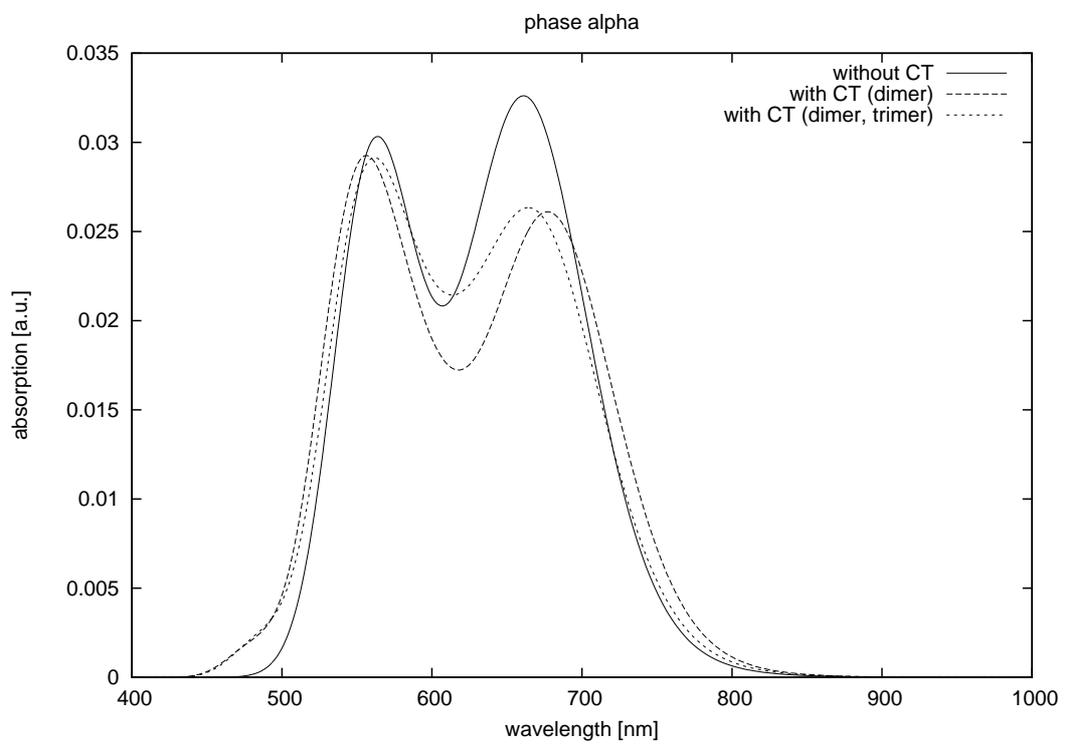}
\caption{Crystal absorption spectrum of TiOPc (phase II).
}
\label{fig:abspec}
\end{figure}

We have considered two excitation configurations,
  homo $\rightarrow$ lumo and homo-1 $\rightarrow$ lumo,  for every CT pair,
 where homo is the highest occupied   molecular orbital 
   and lumo is the lowest  unoccupied molecular orbital.
We have construct adjacency matrix $\mathbf{A}$ for LE-LE interaction,
 and for LE-CT and CT-CT interactions
 based on three CT pairs shown in Fig\ref{fig:hexlattice}.
Then, all the dimers and trimers have been counted using 
 the matrices $\mathbf{A}$ and $\mathbf{A^{(2)}}$ (Eq.\eqref{eq:secpowerofA}).
Number of molecules is 193,
 and number of dimers and trimers we have counted is 477 and 797, respectively.
To calculate the transition dipole moment and electron density,
 we have used density functional theory at the 
 B3LYP/6-31g* level.
The LE-CT and CT-CT  interactions have been calculated
 based on configuration interaction picture (eq.\eqref{eq:LECTint}-\eqref{eq:CTCTtrimer}).
The result of the absorption spectrum of phase II TiOPc is presented in Fig.\ref{fig:abspec}.
In the figure, three specra have been shown,
 first is the spectrum  contain LE-LE interactions only,
 second is that contain LE-LE, LE-CT and CT-CT correspond to dimers (eq.\eqref{eq:CTCTdimer}),
 and last is that contain LE-LE, LE-CT and CT-CT
 correspond to dimers and trimers (eqs.\eqref{eq:CTCTdimer} and \eqref{eq:CTCTtrimer}).
A detailed results of crystal excitaion spectrum will be presented in another report.

\section{Conclusion}\label{sec:conclusion}
We have applied graph theory 
 to the estimation of molecular intergrals including CT.
The graphs for LE-LE interaction and interactions related to CT are
 defined based on the nature of each excitation, respectively.
The graph for interactions related to CT is defined considering
 the $\pi$ orbital overlap  between molecular pairs (dimers).
The adjacency matrix and related quantity are utilized,
 and the trimers we are interested in have been collected systematically.
%

We have demonstrated the absorption spectrum of alpha phase of the TiOPc crystal
 as an example of the applied approach to the molecular crystal.
Physical properties of condensed phase, such as crystal, amorphous, liquid and so on,
 depend on not only intramolecular properties but the intermolecular interactions.
The approach based on the graph theory is a systematic one 
 to manage intermolecular interactions  or connections in condensed phase,
 and can be used other type of interactions,
 such as CH-$\pi$ and intermolecular hydrogen bonds.

\section{Acknowledgement}
We thank to K. Nakai and N. Kobayashi for fruitful discussion
 about TiOPc crystals and their spectra.
We also thank to M. Okuda and S. Matsutani for
 giving encouraging advices and profitable suggestions.

\bibliography{../bibdataofHN}


\clearpage

\clearpage

\clearpage

\clearpage

\end{document}